\documentclass[a4paper,fleqn,usenatbib]{mnras}

\usepackage{graphicx}
\usepackage{amsmath}
\newcommand{\bru}{{\sc{bats-r-us }}}

\newcommand{\tb}{$\tau$ Boo }
\newcommand{\tbb}{$\tau$ Boo}

\newcommand{\Msol}{M_{\odot}}

\title[Temporal variability of the wind from star $\tau$ Boo]{Temporal variability of the wind from the star $\tau$ Bo\"otis}
\author[B. A. Nicholson et al.]{B. A. Nicholson,$^{1,2}$  A. A. Vidotto,$^{3,4}$ M. Mengel,$^{1}$ L. Brookshaw,$^{1}$ B. Carter,$^{1}$ \and P. Petit,$^{5,6}$ S. C. Marsden,$^{1}$ S. V. Jeffers$^{7}$, R. Fares$^{8}$, and the BCool Collaboration\\
$^{1}$Computational Engineering and Science Research Centre, University of Southern Queensland, Toowoomba, Australia\\
$^{2}$European Southern Observatory, Karl Schwarzschild Str. 2, 85748 Garching, Germany\\
$^{3}$Observatoire de l'Universit\'e de Gen\`eve, Chemin des Maillettes 51, Versoix, CH-1290, Switzerland\\
$^{4}$School of Physics, Trinity College Dublin, Dublin-2, Ireland\\
$^{5}$CNRS, Institut de Recherche en Astrophysique et Planétologie, 14 Avenue Edouard Belin, F-31400 Toulouse, France\\
$^{6}$Université de Toulouse, UPS-OMP, Institut de Recherche en Astrophysique et Planétologie, Toulouse, France\\
$^{7}$Institut f\"ur Astrophysik, Universit\"at G\"ottingen, Friedrich Hund Platz 1, 37077 G\"ottingen, Germany\\
$^{8}$INAF- Osservatorio Astrofisico di Catania, Via Santa Sofia, 78, 95123 Catania, Italy}

\begin{document}

\date{}

\pagerange{\pageref{firstpage}--\pageref{lastpage}} \pubyear{2016}

\maketitle

\label{firstpage}

\begin{abstract}
We present new wind models for $\tau$ Bo\"otis (\tbb), a hot-Jupiter-host-star whose observable magnetic cycles makes it a uniquely useful target for our goal of monitoring the temporal variability of stellar winds and their exoplanetary impacts. Using spectropolarimetric observations from May 2009 to January 2015, the most extensive information of this type yet available, to reconstruct the stellar magnetic field, we produce multiple 3D magnetohydrodynamic stellar wind models. Our results show that characteristic changes in the large-scale magnetic field as the star undergoes magnetic cycles produce changes in the wind properties, both globally and locally at the position of the orbiting planet. Whilst the mass loss rate of the star varies by only a minimal amount ($\sim 4$ percent), the rates of angular momentum loss and associated spin-down timescales are seen to vary widely (up to $\sim 140$ percent), findings consistent with and extending previous research. In addition, we find that temporal variation in the global wind is governed mainly by changes in total magnetic flux rather than changes in wind plasma properties. The magnetic pressure varies with time and location and dominates the stellar wind pressure at the planetary orbit. By assuming a Jovian planetary magnetic field for \tb b, we nevertheless conclude that the planetary magnetosphere can remain stable in size for all observed stellar cycle epochs, despite significant changes in the stellar field and the resulting local space weather environment.
\end{abstract}


\section{Introduction}

The study of stellar winds gives insight into the evolution of stars and the planets that orbit them. The wind affects the stellar rotation rate through mass loss and magnetic braking \citep{Schatzman1962,Weber1967,Bouvier2013}, and can also impact the atmospheres of orbiting planets \citep{Adams2011,Lammer2012}, with potential implications on planet habitability \citep{Horner2010,Vidotto2013}. An example of this impact is seen in our own Solar system with the stripping of Mars' atmosphere by the young Sun \citep{Lundin2007}. However, studying stellar winds is problematic, as winds of stars like our Sun are too diffuse to observe directly \citep{Wood2015}. Investigation of these stars therefore requires a combination of observation and magnetohydrodynamic (MHD) simulation in order to estimate the properties of these winds. 

The star $\tau$ Bo\"otis (\tbb, spectral type F7V) is an ideal candidate for a study of the temporal variability of stellar winds, and their potential impacts on orbiting planets. It is the only star to date, other than our Sun, observed to have a magnetic field cycle, with multiple field reversals observed \citep{Catala2007,Donati2008a,Fares2009,Fares2013,Mengel2016}. These observations reveal that the star has a magnetic cyclic period estimated to be a rapid 740 days \citep{Fares2013}. The availability of multiple epochs of magnetic field observations allows us to make observationally informed MHD models of \tbb's wind, such as those detailed by \citet{Vidotto2012} (4 epochs observed between June 2006 and July 2008). 

In this paper we present an additional eight MHD simulations of the winds of \tb from magnetic field observations taken between May 2009 and January 2015. Our aim is to examine the changes in the wind behaviour due to the variations in the magnetic field over the observed epochs, which include multiple polarity reversals, and to demonstrate how these changes could impact the planet \tb b. 

In section \ref{sec:model} we detail the wind model and magnetic field input. The results of our simulations are presented in two parts: Section \ref{sec:global_results} presents the global wind properties over the eight epochs, and in Section \ref{sec:planet_wind} we investigate the properties of the wind around \tb b and the potential impact of the wind on the planet. We discuss these results in Section \ref{sec:Discussion}, make conclusions based on our findings in Section \ref{sec:concl}.

\section[]{Stellar Wind Model}
\label{sec:model}

\subsection{The \bru Code}

The stellar wind model used here is the same as used in \citet{Vidotto2012}, but with higher resolution \citep[as in ][]{Vidotto2014c,Vidotto2015}.  For simulating the winds we use the \bru code \citep{Powell1999,Toth2012}; a three-dimensional code that solves the ideal magnetohydrodynamic (MHD) equations
\begin{equation}
\frac{\partial \rho}{\partial t} + \nabla \cdot (\rho {\mathbf u}) = 0,
\end{equation}
\begin{equation}
\frac{\partial (\rho {\mathbf u})}{\partial t} + \nabla \cdot \left[ \rho {\mathbf u} \otimes {\mathbf u} + \left(P+\frac{B^2}{8\pi}\right)I - \frac{{\mathbf B} \otimes {\mathbf B}}{4\pi} \right]= \rho g,
\end{equation}
\begin{equation}
\frac{\partial {\mathbf B}}{\partial t} + \nabla \cdot \left({\mathbf u} \otimes {\mathbf B} - {\mathbf B}\right) = 0,
\end{equation}
\begin{equation}
\frac{\partial \epsilon}{\partial t} + \nabla \cdot \left[ {\mathbf u} \left( \epsilon + P + \frac{B^2}{8\pi} \right) - \frac{({\mathbf u} \cdot {\mathbf B}){\mathbf B}}{4\pi}\right] = \rho {\mathbf g} \cdot {\mathbf u},
\end{equation}
where $\rho$ and $\mathbf{u}$ are the plasma mass density and velocity, $P$ is the gas pressure, ${\mathbf B}$ the magnetic flux, and ${\mathbf g}$ is the gravitational acceleration due to a star of mass $M_*$. The total energy density, $\epsilon$, is given by
\begin{equation}
\epsilon = \frac{\rho u^2}{2} + \frac{P}{\gamma -1} + \frac{B^2}{8 \pi}.
\end{equation}
The polytropic index, $\gamma$, is defined such that 
\begin{equation}
	P \propto \rho^{\gamma}.
	\label{eq:PpropGamma}
\end{equation}
Table \ref{tab:windinput} lists the stellar and wind values used for this simulation. The values of \tbb's mass and radius, $M_*$ and $R_*$, are taken from \citet{Takeda2007}, and the rotation period, $t_{rot}$, is taken from \citet{Fares2009}. The wind mean particle mass, $\mu$, is chosen on the assumption that the wind is only composed of protons and electrons. Each simulation is initialised using a polytropic wind solution and the magnetic field information from observations of a given epoch. It is then iterated forward around 30,000 time steps. This ensures that a steady state is achieved. Each simulation, therefore, is a snapshot of the wind at each epoch. 
\begin{table}
\centering
\caption{Adopted stellar parameters for \tb used in the wind model. }
\label{tab:windinput}
\begin{tabular}{lc}
\hline
$M_*$ ($M_{\odot}$) & 1.34 \\
$R_*$ ($M_{\odot}$) & 1.46 \\
Rotation Period, $t_{rot}$ (days) & 3.0 \\
Base Wind Temperature (K) & $2 \times 10^6$\\
Base Wind Density (g/cm$^3$) & $8.36 \times 10^{-16}$\\
Wind Mean Particle Mass, $\mu$ & 0.5 \\
Polytropic Index, $\gamma$ & 1.1 \\
\hline
\end{tabular}
\end{table}

\subsection{Adopted surface magnetic fields}

The \bru code uses a star's surface magnetic field as input to the inner boundary conditions. The magnetic field information for \tb comes from spectropolarimetric observations with the NARVAL instrument at T\'elescope Bernard Lyot in the Midi Pyr\'en\'ees, accessed through the BCool collaboration \citep{Marsden2014}. The polarization information in the spectra observed at different stellar rotation phases are used to reconstruct the surface magnetic field topologies using Zeeman Doppler Imaging \citep[ZDI,][]{DonatiLandstreet2009}. The reconstruction of the magnetic field maps used in this work is presented in \cite{Mengel2016}. The radial field topologies are of sole interest for the stellar wind modelling as the other, non-potential field components have been shown to have a negligible effect on the wind solution \citep{Jardine2013}. 

\begin{figure*}
\includegraphics[width=\textwidth]{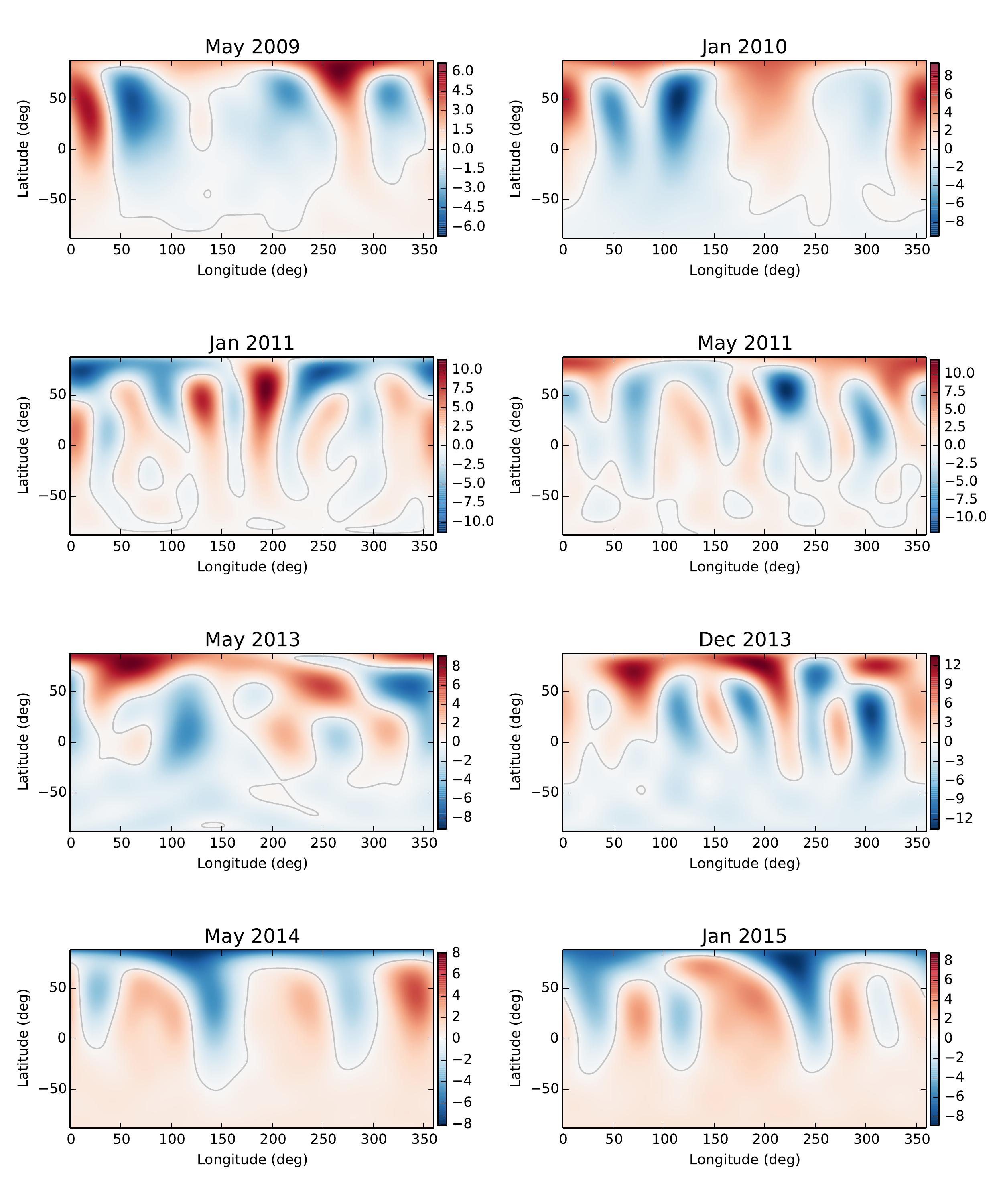}
\caption{Radial magnetic field maps for the eight observed epochs, measured in Gauss (G), with the grey line indicating $B_r=0 G$.  The May 2009 to January 2011 maps come from \citet{Fares2013}, and the May 2011 to January 2015 maps from \citet{Mengel2016}. The Fares et al. epochs have been reanalysed for consistency with the Mengel et al. maps. The field remains dominantly poloidal across the eight observations, and three polarity reversals are observed. It is believed that two polarity reversals have occurred between May 2011 and May 2013.  }
\label{fig:magfields}
\end{figure*}

Figure \ref{fig:magfields} shows the reconstructed radial magnetic fields, in units of Gauss (G), for eight sets of observations (epochs): May 2009, January 2010, January 2011, May 2011, May 2013, December 2013, May 2014 and January 2015. The grey lines here are contours at 0 G. The first three of these epochs were published by \citet{Fares2013}; however, for consistency the maps presented here are an updated version using the same methodology as used with the other five epochs, the details of which are presented in \cite{Mengel2016}. Due to the inclination of the star, the area below $-40^\circ$ is not observable. The field in this area of the stellar surface is solved for by enforcing $\nabla \cdot {\mathbf B} = 0$. \cite{Vidotto2012} showed that choosing this constraint, versus explicitly forcing a symmetric or antisymmetric solution to the large-scale field topology has little impact on the wind solution, especially in the visible hemisphere. 

Table \ref{tab:mag} shows a summary of the global magnetic field properties over the observed epochs. The magnetic cycle phase is calculated based on a 740-day cycle found by \citet{Fares2013}, with the cycle zero phase chosen to be 2453818 Heliocentric Julian Date (HJD) to align with the magnetic cycle zero-point of \citet{Vidotto2012}. The complexity of the field topology can be quantified by examining the amounts of energy in the different set spherical harmonics that are used to describe the field \citep{Donati2006}. The modes of these coefficients where $l \leq 2$ give the dipolar and quadrupolar field configurations. The complexity of the large-scale field is quantified by the percentage of magnetic energy present in the modes where $l > 2$ over the total energy. A lower percentage indicates a simpler field configuration, whereas a higher percentage means that a larger amount of energy is present in more complicated, smaller scale features. 

The field remains dominantly poloidal throughout all observations, with most of the magnetic energy being contained in more complex field components and the complexity varying from 45 percent to 83 percent over the cycle. The star is observed to undergo three polarity reversals, with two reversals assumed to have occurred between the May 2011 and May 2013 observations. The absolute surface radial magnetic flux (see Section \ref{sec:derived_wind}) ranges from $1.53 \times 10^{23}$ Maxwells (Mx) to a peak of $3.28 \times 10^{23}$ Mx before the polarity reversal that was then seen in the May 2014 observations. This same peak and fall in magnetic flux is not observed in the previous polarity reversals probably due to the timing of the observation, and the length of time between observed epochs. 

\begin{table}
\centering	
\caption{Summary of the radial magnetic field polarity and complexity of $\tau$ Boo for observations from May 2009 to January 2015. The magnetic cycle period is taken to be 740 days from \citet{Fares2013}, with the phase zero point at 2453818 HJD. }
\label{tab:mag} 
\begin{tabular}{ccccc}
\hline
Date & Magnetic  & Visible Pole & Radial Field\\
     & Cycle Phase  & Polarity  &  Complexity \\
	 &				&			& ($\%$ of $l > 2$)\\
\hline
May 2009 & 0.57 & Positive & 77 \\

Jan 2010 & 0.91 & Positive & 45 \\

Jan 2011 & 0.38 & Negative & 73 \\

May 2011 & 0.54 & Positive & 83 \\

May 2013 & 0.51 & Positive & 62 \\

Dec 2013 & 0.81 & Positive & 65 \\

May 2014 & 0.02 & Negative & 58 \\

Jan 2015 & 0.34 & Negative & 62 \\
\hline
\end{tabular}
\end{table}

\section[]{Wind Simulation Results: Global Wind Properties}
\label{sec:global_results}
\subsection{Magnetic Field}

The behaviour of the stellar wind is dependent on the geometry and strength of the global stellar magnetic field. Here we examine the behaviour of the wind at intervals of five months to one year, over a total period of five years and seven months. It is expected that the wind will vary little over the time it takes to observe one epoch \citep[$\sim 14$ days][]{Vidotto2012}.  

\begin{figure*}
\includegraphics[width=0.90\textwidth]{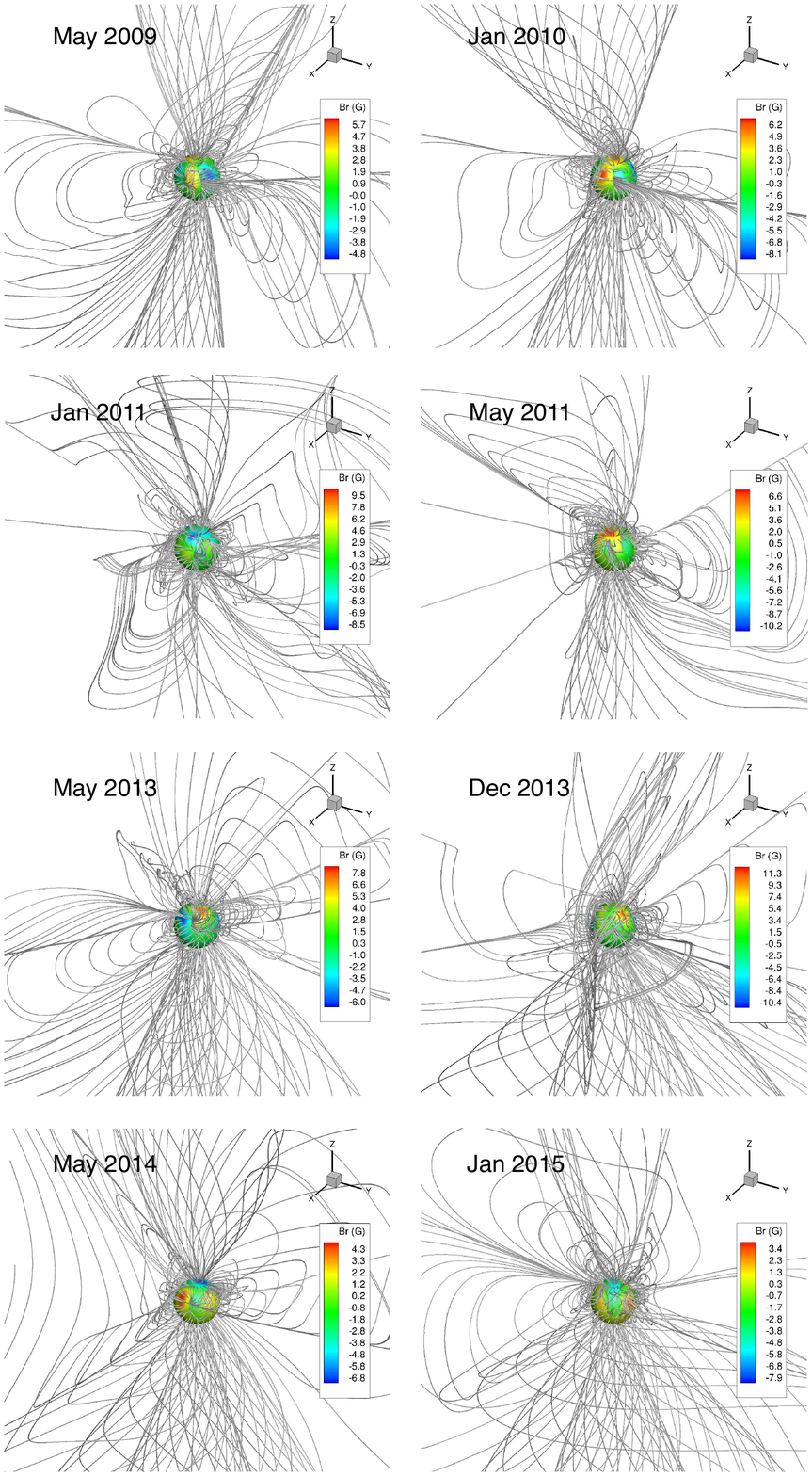}
\caption{These plots show the simulation results of the large-scale magnetic field lines surrounding \tbb. The sphere in the centre represents the stellar surface, with the colour contours indicating the radial magnetic field strength at the surface. The rotational axis of the star is along the $z$-axis, with the equator lying in the $xy$-plane. It can be seen that the field lines become twisted around the axis of rotation ($z$-axis) due to the presence of the wind. }
\label{fig:sim3dmagfields}
\end{figure*}

As the magnetic field extends outward from the stellar surface, it will be influenced by the presence of the stellar wind. Figure \ref{fig:sim3dmagfields} shows the magnetic field lines (grey lines) outwards from the surface of the star. The colour contours on the surface represent radial magnetic flux. The lines are seen to twist along the rotation axis ($z$-axis) of the star. The number of large closed loops is notable compared to similar plots produced by \citet{Vidotto2012}, which is due to a difference in simulation resolution and different reconstruction process to create the magnetic maps. The impact of grid resolution on results is explored in more detail in Appendix \ref{sec:grid_size}. 

This proportion of open to closed magnetic flux can be quantified by examining magnetic flux at different points in the simulation. The unsigned radial flux, $\Pi$, is given by
\begin{equation}
	\Pi = \oint_{S_R} | B_r | dS_R,
	\label{eq:Pi}
\end{equation} 
where $S_R$ is a spherical surface of radius $R$. The unsigned surface radial magnetic flux, $\Pi_0$, can be calculated from Equation \ref{eq:Pi} at the surface $R=R_*$. At $R\geq10R_*$ the magnetic flux is contained solely within open field lines, and Equation \ref{eq:Pi} integrated over a sphere $S_{R\geq10}$ in this region then represents the absolute open magnetic flux, $\Pi_{open}$. The fraction of open flux, $f_{open}$ is then defined as 
\begin{equation}
	f_{open} = \frac{\Pi_{open}}{\Pi_0}. 
\end{equation}
These values are given in Table \ref{tab:wind}. The proportion of open to closed flux is seen to be low, with a majority of the flux from the surface being contained within closed field lines. 

\subsection{Derived Wind Properties}
\label{sec:derived_wind}

Quantities of interest to the study of stellar rotation evolution, such as mass loss, angular momentum loss and spin-down timescale, can be calculated from the output of the wind simulation. Table \ref{tab:wind} summarises the properties of the wind derived from our simulation.  The mass loss rate,
\begin{equation}
	\dot{M} = \oint_{S_{R\geq10}} \rho \mathbf{u} \cdot dS_{R\geq10},
\end{equation}
and angular momentum loss rate
\begin{equation}
	\dot{J} = \oint_{S_{R\geq10}} \left[ \frac{ - B_{\phi} B_r \sqrt{x^2+y^2} }{ 4 \pi} + u_\phi \rho u_r \sqrt{x^2+y^2} \right] dS_{R\geq10}
\end{equation}
\citep{Mestel1999,Vidotto2014c} are also evaluated over $S_{R\geq10}$, where these quantities reach a constant value. The angular momentum loss is used to infer the timescale of magnetic braking, $\tau$, defined as $\tau = J/\dot{J}$, measured in Gyr, where J is the angular momentum of the star given by $J=(I_{core} + I_{envelope})\Omega_*$, with $\Omega_*$ being the stellar angular velocity. We estimate the spin-down time of \tb by using stellar evolution models to estimate the moment of inertia of the core, $I_{core}$, and convective envelope, $I_{envelope}$, separately. This spin-down time-scale, $\tau$, is given by 
\begin{equation}
	\tau = \frac{2 \pi (I_{core} + I_{envelope})}{t_{rot} \dot{J}}.
\end{equation}
Table \ref{tab:wind} gives this these spin-down times using $I_{core} = 1.05 \times 10^{54}$ g cm$^2$ and $I_{envelope} = 4.53 \times 10^{51}$ g cm$^2$, calculated from the model of \cite{Baraffe1998} (Gallet, private communication). 
\begin{table*}
\centering
\begin{minipage}{140mm}
\centering	
\caption{Summary of the simulated global wind properties of $\tau$ Boo based on observations from May 2009 to January 2015. The behaviour of the magnetic field is described by the unsigned surface flux, $\Pi_0$, the open flux beyond 10 stellar radii, $\Pi_{open}$, and the ratio of these quantities, $f_{open}$. These values indicate a significant variation in magnetic field behaviour between epochs. The stellar mass loss rate, $\dot{M}$, varies an insignificant amount over the observed epochs. The angular momentum loss rate, $\dot{J}$, however, and associated spin down timescales, $\tau$, are observed to vary significantly over the eight epochs, correlating with the observed changes in the magnetic field.  }
\label{tab:wind}
\begin{tabular}{c c c c c c c}
\hline
Date   & $\Pi_{0}$ 	   & $\Pi_{open}$  & $f_{open}$ & $\dot{M}$ 				   & $\dot{J}$ & $\tau$ \\
       & ($10^{22}$Mx) & ($10^{22}$Mx) & 			& ($10^{-12}M_{\odot}$ yr$^{-1}$) & ($10^{32}$ergs) & (Gyr)\\
\hline
May 2009 & 15.3 & 4.5 & 0.29 & 2.34 & 1.3 & 6.5\\

Jan 2010 & 22.3 & 7.6 & 0.34 & 2.31 & 2.0 & 4.1\\

Jan 2011 & 22.3 & 5.6 & 0.25 & 2.34 & 1.6 & 5.0\\

May 2011 & 20.0 & 4.4 & 0.22 & 2.31 & 1.3 & 6.1\\

May 2013 & 21.4 & 7.1 & 0.33 & 2.29 & 1.7 & 4.7\\

Dec 2013 & 32.8 & 10.5 & 0.32 & 2.38 & 3.0 & 2.7\\

May 2014 & 18.7 & 8.0 & 0.43 & 2.30 & 1.6 & 5.2\\

Jan 2015 & 21.6 & 8.2 & 0.38 & 2.28 & 1.7 & 4.5\\
\hline
\end{tabular}
\end{minipage}
\end{table*}

The mass loss rates show little variation over the eight epochs, ranging between $2.29 \times 10^{-12}$ to $2.38 \times 10^{-12}$ $\Msol $yr$^{-1}$, approximately 100 times the Solar mass loss rate.  This level of variation ($\sim 4$ percent) is in agreement with the epoch-to-epoch mass loss rate variability found by \citet{Vidotto2012} ($\sim 3$ percent). 

We calculate a lower limit on the x-ray luminosity, as in \cite{Llama2013}, assuming that the quiescent x-ray emission of the coronal wind is caused by free-free radiation. We find little variation ($\sim 2\%$) over the observed epochs. This is in agreement with the previous wind model results of \cite{Vidotto2012}, and the x-ray observations of \tb by \cite{Poppenhaeger2012} and \cite{Poppenhaeger2014}.

The angular momentum loss rate values are seen to vary by $ \sim 140 $ percent over the eight epochs, with a peak at December 2013, corresponding to a peak in the observed radial surface flux, $\Pi_0$, as do the spin-down timescales, $\tau$. Our values of spin down time differ by 1 order of magnitude than those derived by \cite{Vidotto2012}. This is due to different assumptions in the moment of inertia of the star (\cite{Vidotto2012} assumed that of a solid sphere, while we use a more sophisticated approach).

\section{Wind environment around the planet}
\label{sec:planet_wind}
\subsection{Wind properties at \tb b}
Since the planet \tb  b is tidally locked to its star in a 1:1 resonance, the location of the planet with respect to the surface of the star does not change. The planet's orbit lies in the equatorial (x-y) plane of the star \citep{Brogi2012} on the negative x-axis at $x=-6.8 R_*$. Table \ref{tab:planet_environment} gives the properties of the environment surrounding the planet. 
The total pressure, $P_{tot}$, is the sum of the thermal, ram and magnetic pressures. The thermal pressure due to the wind, $P$, is an output of the simulation. The ram pressure, $P_{ram}$, is given by:
\begin{equation}
	P_{ram} = \rho |\Delta \mathbf{u}|^2
\end{equation}
where $|\Delta \mathbf{u}| = |\mathbf{u} - \mathbf{v_k}|$ is the relative velocity between the planet and the wind, with $\mathbf{v_k}$ being the planet's Keplerian velocity. The magnetic pressure, $P_{mag}$ is given by:
\begin{equation}
	P_{mag} = \frac{|{\mathbf B}|^2}{8 \pi}.
\end{equation}
The variation in surface absolute magnetic flux shown in Table \ref{tab:wind} is reflected in the variation in level of absolute magnetic flux, $|\mathbf{B}|$, at the position of the planet. In contrast to this, the wind velocity, $|\mathbf{u}|$, and particle density, $\Gamma$, vary only a small amount ($\sim17$ percent and $\sim14$ percent, respectively) over the observed epochs. The temperature, however, is seen to change by nearly approximately $46$ percent, and total pressure, $P_{tot}$, varies between maxima and minima by $\sim 94$ percent.

\begin{table*}
\centering
\begin{minipage}{140mm}
\centering	
\caption{Summary of the wind properties at the position of the planet $\tau$ Boo b. The variations in absolute magnetic field, $|\mathbf{B}|$, reflects the variation in the observed field. The influence of the magnetic flux changes are seen in variations of the total pressure experienced by the planet.  }
\label{tab:planet_environment}
\begin{tabular}{c c c c c c}
\hline
Date   & $|\mathbf{B}|$ & $|\mathbf{u}|$   & T  & $\Gamma$  & $P_{tot}$ \\
       & ($\times 10^{-2}$ G) & (km s$^{-1}$)& ($\times 10^6$ K) & ($\times 10^6$ Particles cm$^{-3}$) & ($\times 10^{-3}$ dyn cm$^{-2}$) \\
\hline
May 2009 & 0.41 & 209 & 1.05 & 1.46 & 0.81 \\

Jan 2010 & 3.12 & 245 & 1.04 & 1.33 & 1.28 \\

Jan 2011 & 3.65 & 229 & 1.07 & 1.43 & 1.44 \\

May 2011 & 2.46 & 215 & 1.05 & 1.42 & 1.06 \\

May 2013 & 3.06 & 226 & 1.05 & 1.40 & 1.19 \\

Dec 2013 & 4.10 & 232 & 1.52 & 1.52 & 1.57 \\

May 2014 & 2.82 & 216 & 1.11 & 1.41 & 1.16 \\

Jan 2015 & 2.10 & 224 & 1.05 & 1.36 & 1.02 \\
\hline
\end{tabular}
\end{minipage}
\end{table*}

\begin{figure}
	\centering
	\includegraphics[width=86mm]{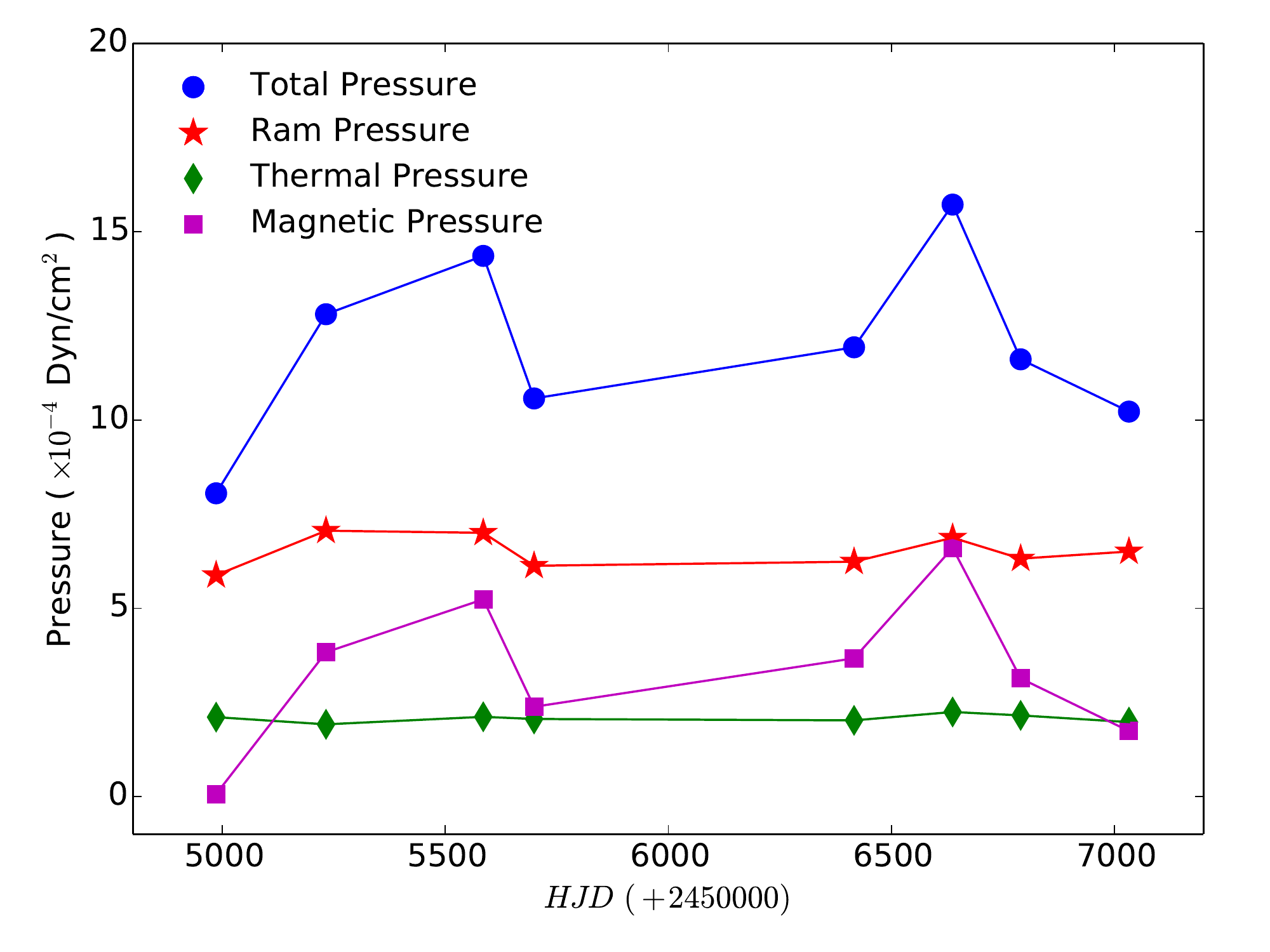}
	\caption{Pressure values at the orbit of the planet \tb b for each observed epoch. The total pressure is the sum of the ram, thermal and magnetic pressures. The ram and thermal pressures vary only slightly over the observed epochs, whereas the magnetic pressure varies more significantly. The lines between points are to guide the eye, and do not represent a fit to the data. }
	\label{fig:pressure}
\end{figure}

Figure \ref{fig:pressure} shows the total, ram, magnetic and thermal pressure measurement for each observed epoch.  The magnetic pressure varies by up to $48$ percent, whereas the ram and thermal pressures varies minimally ($ \sim 5$ percent) over the eight epochs. This suggests that the dominant contributing factor to changes in the total pressure come from the changes in the magnetic pressure.

\subsection{Planetary magnetospheric behaviour}

The external pressure around the planet can be used to infer possible behaviours of the planet's magnetosphere. The ratio of the planetary magnetospheric radius, $R_m$, to the planetary radius, $R_p$, is derived from the equilibrium between the pressure from the wind and the outward force of the planetary magnetosphere. This can be written as
\begin{equation}
	\frac{R_m}{R_p}= \left( \frac{(B_p/2)^2}{8 \pi P_{tot}} \right)^{\frac{1}{6}}
\end{equation}
\citep{Vidotto2012},where $B_p$ is the strength of the planetary magnetic field at the pole. Since there have been no measurements of the magnetic field strength of hot Jupiters to date, we have assumed a planetary magnetic field strength similar to that of Jupiter, which is a maximum of $\sim 14$ G \citep{Bagenal1992}.  The values of $R_m/R_p$ evaluated over the eight epochs are shown in Table \ref{tab:planet}.  

The behaviour of the planetary magnetosphere can also be described in terms of the auroral ring: the area around the poles of the planet over which the planetary magnetic flux is open, allowing the flow of particles to and from the planetary atmosphere. This can be described in terms of the percentage of the planet's surface area: 
\begin{equation}
	A_{open} = (1-\cos(\alpha)) \times 100, 
\end{equation}
where $\alpha$ is the auroral angular aperture, defined as
\begin{equation}
	\alpha = \arcsin \left( \left( \frac{R_m}{R_p} \right)^{-\frac{1}{2}} \right)
\end{equation}
\citep{Vidotto2015}. 
\begin{table}
\centering
\caption{Values for the ratio of planet magnetospheric radius to planet radius, $R_m/R_p$, auroral aperture $\alpha$ and the percentage coverage of the polar cap, $A_{open}$, for the planet $\tau$ Boo b. These have been calculated assuming magnetic field strength at the pole of 14 G. Despite notable changes in the behaviour of the star's magnetic field and in the total pressure exerted on the planet by the stellar wind, these values remain quite stable over the observed epochs. }
\label{tab:planet}
\begin{tabular}{c c c c }
\hline
Date   & $R_m/R_p$ & $\alpha$  & $A_{open}$ \\
       &           & (degrees)   & ($\%$) \\
\hline
May 2009 & 3.67 & 31.5 & 14.7 \\

Jan 2010 & 3.57 & 31.9 & 15.1 \\

Jan 2011 & 3.56 & 32.0 & 15.2 \\

May 2011 & 3.64 & 31.6 & 14.8 \\

May 2013 & 3.62 & 31.7 & 14.9 \\

Dec 2013 & 3.55 & 32.1 & 15.3 \\

May 2014 & 3.61 & 31.7 & 15.0 \\

Jan 2015 & 3.62 & 31.7 & 14.9 \\
\hline
\end{tabular}
\end{table}

Despite noticeable changes in the total pressure exerted in the planet by the star, the planet's magnetosphere remains around 3.6 times the planet's radius, varying by only $\sim 3$ percent over the observed epochs. This is due to the relative insensitivity of $R_m/R_p$ to changes in $P_{tot}$ (due to the $-1/6$ power dependency). As with the planetary magnetospheric radius, there are minimal changes in the auroral aperture and polar surface area, which remain around $32^{\circ}$ and $15$ percent, respectively. From this we can infer that despite considerable changes in the behaviour of \tbb's magnetic field, \tb b's magnetosphere remains relatively stable.

\section{Discussion}
\label{sec:Discussion}

\subsection{Global wind properties}

The wind simulations presented here are an advance on models that do not use observationally reconstructed magnetic fields as input, or that assume a simplified stellar magnetic field topology. The coronal base temperature and density are poorly constrained by observations, and are free parameters of our model. We chose a wind base temperature that is typical of stellar coronae, and the same base density as adopted in \cite{Vidotto2012}. Out estimated mass loss rate ($\sim 2.3 \times 10^{-12}M_{\odot}$ yr$^{-1}$)is within the range of previous estimates of $1.67 \times 10^{-12}M_{\odot}$ yr$^{-1}$ \citep{Stevens2005} to $6.6 \times 10^{-12}M_{\odot}$ yr$^{-1}$ \citep{Cranmer2011}. Given that Sun's corona is adequately described by the adiabatic process given by equation \ref{eq:PpropGamma}, with the index $\gamma = 1.1$ \citep{VanDoorsselaere2011}, we assume the same for this star.

The behaviour of the angular momentum loss rate and associated spin-down time-scales found here agree with the predictions of \cite{Gallet2013}, who computed rotational evolution models based on wind-breaking laws derived for magnetised Solar-type stars. They find that the spin down timescales of stars at 1 Gyr old, which is the approximate age of \tb \citep{Borsa2015}, should converge to the length of a few Gyrs, the same as presented in the results here. This further strengthens our choice of model parameters. 

The lifetime of a main-sequence star, $\tau_{MS}$, with the mass of \tb (1.34 $M_{\odot}$) is expected to be $\sim 4.8$ Gyrs ($\tau_{MS} = 10(M_*/M_{\odot})^{-2.5}$). Given \tbb’s estimated age of ~1Gyr and our calculated mean spin-down time of $\sim4.9$ Gyr, this implies that Tau Boo will remain a rapid rotator throughout its main sequence lifetime, provided that only the stellar wind is affecting its rotational evolution.

The changing stellar magnetic field polarity of the poles does not have an effect on the wind solution. This is because there is no preferred ‘up’ or ‘down’ orientation of the star, and the global wind properties are calculated as surface integrals around the star. Instead, it is the changing field strength as the star undergoes its reversals that drives changes in the wind behaviour. 

Over the cycle there is a change in field complexity, and this is anti-correlated with on the fraction of open flux $f_{open}$ (linear  correlation coefficient = -0.72). As the field becomes more complex, the amount of flux contained in closed lines increases and the fraction of open flux decreases \citep[see also][]{Lang2012}. No correlation is found, however, between changes in field complexity and changes, or lack there of, in the angular momentum or mass loss rates.

\subsection{Wind-planet interaction}

Wind simulations can give insight into the potential behaviour of the planetary magnetosphere in the presence of the stellar wind. Since there have been no observations of magnetic fields of exoplanets to date, assumptions must be made as to what magnetic field can be expected from \tb b. There is much discussion over the possible magnetic fields of exoplanets. It has been theorised that close in hot-Jupiters such as \tb b are thought to have a weaker magnetic field than similar planets further away from the host star due to the tidal locking slowing the planet's rotation, and hence reducing its magnetic field \citep{Griessmeier2004}. However, there are some studies that indicate that planetary rotation does not directly influence the strength of a planet’s magnetic field \citep{Christensen2009}, but plays a role in the field geometry \citep{Zuluaga2012}. Given the uncertainty in hot-jupiters magnetic fields and the similar nature of this planet to Jupiter, we have assumed a Jovian planetary magnetic field strength for this work. The resulting magnetosphere suggests the planet is protected from the stellar wind, despite large changes in the magnetic field and wind strengths over the cycle. 

The minimum field strength required to sustain a magnetosphere above the surface of \tb b (i.e. $R_m > R_p$) can be estimated by examining the condition of the space weather environment during at the most extreme part of the magnetic cycle (i.e, when $P_{tot}$ is at a maximum). This is calculated to be $\sim 0.4$ G. This does not mean, however, that the planet is protected at this point, as at this limit the Auroral aperture reaches 90 degrees, exposing $100$ percent of the planets surface area to particle inflow and outflow. If we were to call a planet `protected' provided less than $25$ percent of its surface area was contained within the auroral aperture, then the minimum magnetic field strength to achieve this is $\sim 4.7$ G for the \tb system.

\section{Summary and conclusions}
\label{sec:concl}
%
This study examines the variability in the wind behaviour of the star \tb over eight epochs from May 2009 to January 2015 - the most extensive monitoring of the wind behaviour of a single object to date (apart from the Sun). The winds are examined globally to study the star's rotational evolution, and locally around \tb b for the possible impacts the wind might have on the planet's magnetosphere. 

Despite significant changes in the magnetic field behaviour, the mass loss rates do not significantly vary from epoch to epoch ($\sim 4 $ percent), remaining around $2.3 \times 10^{-12}$ $\Msol $yr$^{-1}$. However, the angular momentum loss rate is observed to change considerably over the eight observations, ranging from $1.3 \times 10^{32}$ ergs to $3.0 \times 10^{32}$ ergs. These findings are consistent with angular momentum loss rates and 
 and associated spin-down time-scales predicted by stellar evolution models. 

Examining the wind environment of the planet shows that variations in the absolute flux due to changes in the magnetic field behaviour of the star are reflected in changes in the local space weather of \tb b. Despite these changes, the magnetosphere from an assumed Jupiter-like planetary magnetic field is relatively invariant over the observed epochs, with the magnetospheric radius remaining around 3.6 times the size of the planetary radius. 

\section*{Acknowledgments}
This work was partly funded through the University of Southern Queensland Strategic Research Fund Starwinds Project. Some of the simulations presented in this paper were computed on the University of Southern Queensland's High Performance Computer. This research has made use of NASA's Astrophysics Data System. This work was carried out using the \bru tools developed at The University of Michigan Center for Space Environment Modeling (CSEM) and made available through the NASA Community Coordinated Modeling Center (CCMC). This work was supported by a grant from the Swiss National Supercomputing Centre (CSCS) under project ID s516. AAV acknowledges support from the Swiss National Science Foundation through an Ambizione Fellowship. Many thanks to Florian Gallet for providing the moment of inertia values for this work, and many thanks also to Colin Folsom for his help and advice. 

\bibliographystyle{mnras}
\bibliography{bibTauBoo}

\appendix

\begin{table*}
\centering
\begin{minipage}{140mm}
\centering
\caption{This table shows the changes in the May 2011 global wind parameters with differing grid refinement levels. These variations are on similar to or greater than the variation between epochs. As such it is important to ensure that the grid levels are the same when comparing global wind properties between simulations. }
\label{tab:grid}
\begin{tabular}{c c c c c c }
\hline
Grid refinement & Total number   & Smallest cell 		  & $\dot{M}$ 			 & $\dot{J}$  	 & $f_{open} $ \\
    level       & of cells &($\times 10^{-2}$ $R_*$)& ($10^{-12} \Msol yr^{-1}$)  & ($10^{32}$ergs)& 		  \\
\hline
8 & $2.78 \times 10^6$& 5.7 & 2.9 & 1.8 & 0.35 \\

9 & $7.10 \times 10^6$ & 2.9 & 2.5 & 1.5 & 0.23 \\

10 & $3.92 \times 10^7$ & 1.4 & 2.3 & 1.3 & 0.22\\

\hline
\end{tabular}
\end{minipage}
\end{table*}
\begin{table*}
\centering
\begin{minipage}{140mm}
\centering
\caption{This table shows the changes in the May 2011 local wind environment around \tb b with differing grid refinement levels. The variation in due to grid refinement level is on the same level or greater than the variation epoch to epoch. As such it is important to ensure that the grid levels are the same when comparing wind models across epochs. }
\label{tab:grid_planet}
\begin{tabular}{c c c c c}
\hline
Grid refinement &  $|\mathbf{u}|$ & $|\mathbf{B}|$		& $P_{tot}$ 					 & $R_m/R_p$ \\
    level       & (km s$^{-1}$)	  & ($\times 10^{-2}$ G)& ($\times 10^{-3}$ dyn cm$^{-2}$) & \\
\hline
8 & 230 & 1.46 & 1.18 & 3.47\\

9 & 221 & 2.15 & 1.08 & 3.58\\

10 & 215 & 2.46 & 1.06 & 3.64\\

\hline
\end{tabular}
\end{minipage}
\end{table*}
\section{Sensitivity of results to grid refinement level}
\label{sec:grid_size}
The design of the \bru code allows the simulation grid to be constructed so that areas of interest can be studied in greater detail by local grid refinement. This means that the region closest to the surface of the star that is changing the greatest can be simulated at a much finer resolution, but leaving the outer regions of the simulation with a much coarser grid structure to save on computing resources. 

This section examines how the level of refinement of the grid, that is, the number of times the grid in the region close to the star is subdivided into smaller cells, affects the simulation outcomes. Using the May 2011 data, we examined $\dot{M}$, $\dot{J}$ and $f_{open}$ (as calculated in Section \ref{sec:derived_wind}) for three different refinement levels. Table \ref{tab:grid} shows these global wind parameters, along with the total number of computational cells and smallest cell size for each refinement level. In our grid design, the grid refinement changes occur in the simulation region close to the surface of the star, so that the grid size out beyond $10R_*$ remains the same. 

The values of the wind parameters are seen to vary across refinement levels, and these variations are on the same order as the variations observed across epochs, or greater as in the case of mass loss rate. However, these variations are much smaller than the observational uncertainties and theoretical limits of both $\dot{M}$ and $\dot{J}$. The changes in $f_{open}$ are not unexpected as the finer cell structure means that more of the magnetic field structure is being resolved, and field lines that could be taken to be open in a coarser grid structure are found to be closed at a higher refinement level. 

Table \ref{tab:grid_planet} shows the changes in the simulation results at the position of \tb b due to changes in grid refinement. As with the global wind properties, the local wind properties at the planet vary between grid sizes on the same scale as variations between epochs. Even though it might appear that an under-resolved grid overestimates the impact of the wind on the planet, these variations are smaller than the observational uncertainties on these parameters. 

Given the variations seen due to different grid sizes it is important to use the same grid refinement level to compare simulations of different epochs. Using a higher resolution would give marginally more accurate results, but given the unreasonable amount for computation requires to reach refinement level 11 (total number of cells $\sim 2.86 \times 10^8$), we conclude that refinement level 10 is the most appropriate for the current study.


\label{lastpage}

\end{document}